\documentclass[sigconf]{acmart}
\usepackage{float}
\usepackage{mathtools}
\usepackage{multirow}
\usepackage{listings}
\usepackage[flushleft]{threeparttable}
\usepackage{afterpage}
\def\BibTeX{{\rm B\kern-.05em{\sc i\kern-.025em b}\kern-.08em
    T\kern-.1667em\lower.7ex\hbox{E}\kern-.125emX}}

 \usepackage{dcolumn}
 
\AtBeginDocument{%
  \providecommand\BibTeX{{%
    \normalfont B\kern-0.5em{\scshape i\kern-0.25em b}\kern-0.8em\TeX}}}

\setcopyright{acmcopyright}
\copyrightyear{2018}
\acmYear{2018}
\acmDOI{10.1145/1122445.1122456}

\acmConference[CHASE 2020]{CHASE 2020: International Workshop on Cooperative and Human Aspects of Software Engineering}{May 24, 2020}{Seoul, South Korea}
\acmBooktitle{CHASE 2020: International Workshop on Cooperative and Human Aspects of Software Engineering, May 24, 2020, Seoul, South Korea}
\acmPrice{15.00}
\acmISBN{978-1-4503-XXXX-X/18/06}

\begin{document}

\title{Multitasking Across Industry Projects: \\A Replication Study}

\author{Karina Kohl}
\email{karina.kohl@edu.pucrs.br}
\orcid{0000-0002-2964-4681}
\affiliation{%
  \institution{School of Technology, PUCRS}
  \country{Brazil}
}

\author{Bogdan Vasilescu}
\email{vasilescu@cmu.edu}
\orcid{0000-0002-2650-6638}
\affiliation{%
  \institution{Carnegie Mellon University}
  \country{USA}
}

\author{Rafael Prikladnicki}
\email{rafael.prikladnicki@pucrs.br}
\orcid{0000-0003-3351-4916}
\affiliation{%
  \institution{School of Technology, PUCRS}
  \country{Brazil}
}

\renewcommand{\shortauthors}{Kohl and Vasilescu, et al.}

\begin{abstract}
  \textit{Background}: Multitasking is usual in software development. It is the ability to stop working on a task, switch to another, and return eventually to the first one, as needed or as scheduled. Multitasking, however, comes at a cognitive cost: frequent context-switches can lead to distraction, sub-standard work, and even greater stress. \textit{Aims}: This paper reports a replication experiment where we gathered data on a group of developers from a software development company from industry on a large collection of projects stored in GitLab repositories. \textit{Method}: We reused the developed models and methods from the original study for measuring the rate and breadth of a developers' context-switching behavior, and we study how context-switching affects their productivity. We applied semi-structured interviews,  replacing the original survey, to some of the developers to understand the reasons for and perceptions of multitasking. \textit{Results}: We found out that industry developers multitask as much as OSS developers focusing more (on fewer projects), and working more repetitively from one day to the next is associated with higher productivity, but there is no effect for higher multitasking. Some commons reasons make them multitask: dependencies, personal interests, and social relationships. \textit{Conclusions}: Short context change, less than three minutes, did not impact results from industry developers; however, more than that, it brings a feeling of left the previous tasks behind.  So, it is proportional to how much context is switched: as bigger the context and bigger the interruption, it is worst to come back.
\end{abstract}


% \ccsdesc[500]{Software and its engineering~Software configuration management and version control systems}


\maketitle

\section{Introduction}\label{sec:intro}

Multitasking is the ability to stop working on a task, switch to another, and return eventually to the first one, as needed or as scheduled. The goal is to optimize human resource allocation, while reprioritizing tasks dynamically \cite{Gonzales2005}. When done well, or at least in a disciplined way, multitasking can yield dividends \cite{Adler2012}. If a task in the queue has a higher priority than the current one, switching them can improve performance. However, multitasking comes at a cost though \cite{Borst2015}. Humans, programmers included, have a certain, limited amount of cognitive flexibility, the mental ability to switch from thinking about one concept to thinking about another. Limitations apply to the number of concepts we can juggle, as well as to the difficulty in switching between them. As we can imagine, reaching our innate limitations can result in decreased performance on all tasks and perhaps even diminished quality. It is unknown how far multitasking can be pushed safely, although some anecdotal evidence is available \cite{Vasilescu2016b}. 

Vasilescu et al.\cite{Vasilescu2016b} say that software developers have long been pushing the limits on multitasking because of the innate modularity of the development process and the independence of module processing (e.g., one can code while tests are in execution). They realized that in open-source software, developers also commonly contribute to multiple projects at the same time, bridging different communities. With the advent of social coding tools like GitHub, this has intensified. It is not uncommon to find prolific developers contributing code to 5-10 GitHub projects in the same week. Also, contributing to as many GitHub projects as possible is an accomplishment, valued by peers and employers alike. There are various reasons why developers are more prolific on GitHub compared to other platforms. The features and usability provided by GitHub play a big role. So, with so many drivers for multitasking, it is easy to see how one could cross the limits from safe, productive multitasking into an overloaded mode, where code output falls, and bugs start to multiply. The question becomes, where are those limits, and what are their determinants? 

Through analysis of longitudinal data, Vasilescu et al. \cite{Vasilescu2016b} investigated how productivity (i.e., outputs produced per unit time) of prolific GitHub programmers is determined by the number of projects they work on, how much they focus on each (relative to the others) and, how diverse the projects they contribute to are in terms of programming languages. Notably, the very platform (GitHub) that has introduced this multi-project multitasking phenomenon also gives all the tools we need to understand when programmers are at risk to approach their limits because of it. 

However, software developers working in the industry, frequently also need to contribute to different projects and repositories to have their tasks done. More than that, they have other factors that force them to multitask and switch between different contexts: project meetings, support to other developers and, tasks blocked due to some impediment, for example. Considering this panorama, we identify an opportunity to compare how multitasking differently impacts social developers at GitHub and tech industry developers.

This paper presents a conceptual replication from Vasilescu et al. \cite{Vasilescu2016b} where we reused the study design and experimental steps provided by them. However, we modified the subject pool: instead of the analysis of the multitasking behavior of developers contributing to Open Source projects using GitHub data, the replication analyzed developers working full time in a technology company using anonymized data from their GitLab plataform. We also varied one aspect of the conditions of the experiment: instead of a survey with prolific developers, we performed semi-structured interviews with few developers of the company studied.

Replications play a crucial role in Empirical Software Engineering by allowing the community to build knowledge about which results or observations hold under which conditions. Therefore, not only can a replication that produces similar results as the original experiment is viewed as successful, but a replication that produces results different from those of the original experiment can also be considered as successful. A replication produces different results is useful because it provides insight to help the community understand why the results were different. Gomez et al~\cite{Gomez2014} argue that experimentation helps to build a reliable base of knowledge, reducing uncertainty about theories, methods, tools, etc. Several replications need to run to strengthen the evidence.

The paper is organized as follows. Section~\ref{sec:original} summarizes information about the original study. Section~\ref{sec:replication} summarizes information about the replication study, and Section~\ref{sec:results} presents the results of the replication study. Section \ref{sec:comparison} discusses and compares the results from original study and their implications. Section~\ref{sec:agenda} discusses the future research agenda. Section ~\ref{sec:threats} presents the identified threats to validity, and Section~\ref{sec:conclusions} concludes the paper.

\section{Information About the Original Study}\label{sec:original}

This study is a replication from Vasilescu et al. \cite{Vasilescu2016b}, so to frame it, we followed a similar approach from the original. In this section, we briefly summarize the information about the original study and, detailed information can be consulted in the original paper. First, they seek a deeper understanding of multitasking and its effects on software developers in their daily routines in industry. Second, having identified the critical multitasking and focus switching effects on programmer productivity, they proceed to investigate how these dimensions interact and which trade-offs between them exist. 

% So, they derived the following two research questions:
% \newline
% \hfill \break
% \noindent\textit{\textbf{RQ1.} What are the trends of, reasons for, and effects of multitasking and focus switching on developer productivity in GitHub?}
% \newline \hfill \break
% \textit{\textbf{RQ2.} Are there limits on multitasking (and what are they) before productivity is impacted?}
 
\subsection{Dataset}

The original study used a GitHub dataset that contains information on prolific developers with active contribution history. The dataset comprises contents of all commits authored by 1,255 developers across 58,092 public repositories accessible on GitHub at the time of mining. Developers in the most active three-quartiles (by the total count of active days) were invited for the user survey. Also, the study conceptualizes as a Project, repositories that are grouped when owned by the same GitHub user or having precisely the same name into sets of repositories, once they observed that in some cases, software projects are organized into multiple separate repositories on GitHub.

\subsection{Multi-project Multitasking Productivity}

The original study presents a model for quantitative analysis, to investigate the relationship between outputs produced per unit time (as a proxy for productivity) and a multitude of factors relating to multitasking and focus switching. The authors stated that the time interval to consider it is an important question when studying multitasking and focus switching. During a more extended period, a developer may experience many focus switches without necessarily also “juggling” many tasks over a shorter interval. The modeling to the interplay between focus switching at \textit{two temporal resolutions, daily and weekly,} and using different dimensions. 

The first dimension is \textit{\textbf{Projects Per Day},}, where multitasking activity is measured using the number of different projects contributed to that day. Contributions are measured by commits. It represents a lower bound on the number of switches per day: a developer contributing to \textit{k} projects in one day has to switch focus at least \textit{k} times that day. At the coarsest temporal resolution (week), it is used the average number of projects per day ($AvgProjectsPerDay$) as an aggregate measure of multitasking (inactive days excluded). $AvgProjectsPerDay$ captures the distinction between developers who, over the course of a week, tend to work on projects sequentially day-to-day, and those who interleave contributions to multiple projects each day, i.e., multitask. 

The second dimension is \textit{\textbf{Weekly Focus},} that is useful to distinguish how evenly developers divide their attention among their projects. This differentiation from projects per day is important because it has been found that more narrowly focused developers have less cognitive burden, resulting in higher productivity and quality \cite{rosen2008, xuan2014}. 

So, the uncertainty in a developer's focus switching behavior in a given week, $S_{Focus}$ is defined using the Teachman/Shannon entropy index, a commonly used diversity measure in many scientific disciplines \cite{Brynjolfsson}\cite{BenbunanFich2011AnEI}\cite{xuan2014}. This measure is defined as 
\begin{equation}
S_{Focus} = - \sum_{i=1}^{N}{p_i}  log_2 {p_i}
\end{equation}

where $p_{i}$ is the fraction of the developer’s commits this week, in project \textit{i}, and $N$ is the total number of projects this week. $S_{Focus}$ ranges between 0 when a developer contributes to a single project that week, and $log_2 {N}$, when a developer contributes equally (i.e., $p_{i}$= $1/N$) to all N projects \cite{Vasilescu2016b}.  

Similarly, a developer’s language entropy $S_{Language}$ is measured, defined analogously over the $L$ different programming languages of the files touched that week: 
\begin{equation}
S_{Language} = - \sum_{i=1}^{L}{p_i}  log_2 {p_i}
\end{equation}

$S_{Language}$ is a proxy for the overall complexity of one’s contributions to different projects: when writing code in multiple programming languages, in addition to switching focus between different projects (which involves restoring project specific contexts), one also must switch focus between different languages (which may involve different skills) 

The third dimension is \textit{\textbf{Day-to-Day Focus}} or, $S_{Switch}$, and it can be seen as a measure of the repetitiveness of one’s focus switches from one day to the next: the lower the value, the less repetitive one’s day-to-day behavior is. The original study considered focus switching as a Markov process where the next state is entirely determined by the current one, and the diversity of a developer’s day-to-day focus switches is measured using Markov entropy, and it defines the measure as:

\begin{equation}
S_{Switch} = - \sum_{i=1}^{N}\bigg[{p_i} \sum_{j\in\pi_i}{p}{(j|i)}log_2 {p_i}{(j|i)} \bigg]
\end{equation}

Where $\pi_i$ is the set of outgoing neighbors of node; $p(j|i)$ is the conditional probability that the developer switches focus from $i$ to $j$.

\section{Information About the Replication}\label{sec:replication}
Instead of the analysis of the development activity of GitHub developers, we analyzed the development activity of industry developers from a company in Brazil using their GitLab data. We combined: (1) an analysis of repository data, to quantitatively examine the effects of multitasking and focus switching on productivity; and (2) we used semi-structured interviews to get qualitative insight into the developers’ perceptions of multitasking, focus switching, and their effects.

\subsection{Research Questions}

The two research questions that guide this study are similar to the original ones, except by, we used data from GitLab in the replication. They follow below: \hfill \break

\noindent\textit{\textbf{RQ1.} What are the trends of, reasons for, and effects of multitasking and focus switching on developer productivity in GitLab?}
\newline \hfill \break
 \textit{\textbf{RQ2.} Are there limits on multitasking (and what are they) before productivity is impacted?}

\subsection{Dataset} 
We used a GitLab dataset which considered projects created from 2017 to 2018 and commits on those project from January 1$^{st}$, 2018 to March 20$^{th}$, 2019. Once we were in a controlled industrial environment, we considered all the developers that did at least one commit during the period. Multiple aliases used by a single developer were resolved by similarity. The dataset contains details about the date, size (files added/removed), and contents of 157,658 commits authored by 1,213 developers across 282 projects and 2,453 private repositories accessible on GitLab at the time of mining. Commit data was obtained using the GitLab APIs through Python scripts. The languages of source code files were also obtained thought GitLab APIs. We also interviewed ten developers from the company we analyzed the GitLab dataset.

\subsection{Semi-Structured Interviews}
Instead of a user survey, we conducted semi-structured interviews to get qualitative insight into the developers' perceptions of multitasking, focus switching, and their effects. Semi-structured interviews are a combination of structured and unstructured interviews \cite{Rubin2005}. Such interviews combine specific questions (to bring forth the foreseen information), and open-ended questions (to elicit unexpected types of information). We interviewed ten developers from the company that we analyzed the GitLab data using the following questions:

\begin{itemize}
 \item \textit{How much multitasking do you do in a day? Min/max (perception)};
 \item \textit{How much do you think it would be ideal? In general: positive or negative? Why?};
 \item \textit{From your perspective, what are the main reasons for multitasking?}; 
 \item \textit{How do you plan work across multiple projects if at all?}; 
 \item \textit{Which switches, if any, could be avoided or differently scheduled  if needed?}; 
 \item \textit{Why cross-project interruptions happen?; What are typical triggers for switching from one task/project to another?}; 
 \item \textit{Your productivity (Increase/Decrease)?}; 
 \item \textit{Your code quality (Increase/Decrease)?}; 
 \item \textit{Your happiness/stress levels  (Increase/Decrease)?}; 
 \item \textit{Which are the benefits/negative aspects you see in multitasking?};
 \item \textit{What are typical resume strategies after returning to a previous task/project?}; 
 \item \textit{How does the length of the switch affect your productivity/code quality? Increase project success? Solve more issues? Feel more productive? Contribute more code overall? Review more pull requests?} 
\end{itemize}

The developers are split into two geographically distributed offices in Brazil. The ten developers interviewed are all from an office with around 40 developers allocated in different business areas and teams. We collected their experiences and perceptions about how multitasking impacts their work through the combination of specific questions. We iterated through the open-ended responses using grounded theory methods \cite{Corbin2016}, to categorize them and identify themes. In the process of coding, we separated the data and conceptualized for data analysis, seeking to define and identify the relationship between them. The interviewed participants reported development experience was 9.2 years on average (min 3; max 15). They were 29.8 years old on average (min 23; max 35).

\begin{table*}[]

\centering \small
\caption{OSS versus Industry}
\label{tbl:ghvsgl}
\begin{tabular}
{
p{0.12\linewidth}
{r}p{0.08\linewidth}{r}p{0.08\linewidth}
{r}p{0.08\linewidth}{r}p{0.08\linewidth}
{r}p{0.08\linewidth}{r}p{0.08\linewidth}
{r}p{0.08\linewidth}{r}p{0.08\linewidth}
{r}p{0.08\linewidth}{r}p{0.08\linewidth}
}

\hline
\multicolumn{1}{c}{Statistic} & \multicolumn{2}{c}{Mean} & \multicolumn{2}{c}{St. Dev.} & \multicolumn{2}{c}{Min} & \multicolumn{2}{c}{Median} & \multicolumn{2}{c}{Max} \\
\hline

\hline
\ & GH & GL & GH & GL & GH & GL & GH & GL & GH & GL \\
\hline

GlobalTime & 1,142.76 & 1489.72 & 74.24 & 17.95 & 991 & 1461.00 & 1,12 & 1490.00 & 1.26  & 1524.00\\

UserTime & 308.05 & 23.53 & 194.12 & 17.01 & 1 & 1.00 & 271 & 20 & 2,31 & 63.00\\

% Repositories & 3.70 & 2.35 & 2.22 & 2.40 & 2 & 1 & 3 & 1 & 18  & 43 \\

Projects    & 2.57 & 2.41 & 0.98 & 0.81 & 2.00 & 2.00 & 2.00 & 2.00 & 9  & 9 \\

DaysActive  & 3.98 & 3.08 & 1.63 & 1.24 & 1 & 1.00 & 4 & 3.00 & 7  & 7.00 \\

Languages & 3.19 & 2.26 & 1.43 & 1.04 & 1 & 1.00 & 3 & 2.00 & 14  & 7.00  \\

\hline

Commits & 23.89 & 14.77 & 30.06 & 15.11 & 2 & 2.00 & 15 & 10.00 & 943  & 345.00 \\

FileTouches & 88.56 & 10136.67 & 174.31 & 50645.56 & 2 & 3.00 & 39 & 862.00 & 2,727  & 1036429.00 \\

LOCAdded    & 4.151.36 & - & 15,691.15 & - & 1 & - & 635 & - & 265,702  & - \\

FilesAdded  & - & 7036.79 & - & 37136.06 & - & 2.00 & - & 544.00 & -  & 807232.00 \\

LOCDeleted  & 2,282.19 & - & 9,217.39 & - & 0 & - & 247 & - & 148,977  & - \\

FilesDeleted & - & 3099.87 & - & 26859.33 & - & 0.00 & - & 148.00 & - & 945503.00 \\

\hline

AvgProjectsPerDay & 1.44 & 1.49 & 0.48 & 0.47 & 1 & 1.00 & 1.33 & 1.40 & 6.00  & 6.00 \\

SFocus & 0.92 & 0.94 & 0.44 & 0.36 & 0.02 & 0.09 & 0.92 & 0.92 & 3.01  & 2.81 \\

SLanguages & 0.82 & 0.49 & 0.52 & 0.51 & 0.00 & 0.00 & 0.84 & 0.31 & 2.92  & 2.27 \\

SSwitch & 0.64 & 0.52 & 0.53 & 0.50 & 0.00 & 0.00 & 0.69 & 0.59 & 2.83  & 2.95 \\
\hline
\end{tabular}
\begin{tablenotes}
      \small
      \item GH (GitHub): Summary statistics for week-level data (78,552 rows; 1,193 developers; outliers removed)
      \item GL (GitLab): Summary statistics for week-level data (11,987 rows; 1,213 developers; outliers removed)
    \end{tablenotes}
\end{table*}

 \subsection{Regression Analysis}

The variability in outputs produced (i.e., $FileTouches$) per unit time, our productivity proxy, was modeled as dependent on control measures and the three dimensions of multitasking: projects per day, weekly focus, and day-to-day focus (as in the original study). For each developer, the data consists of measurements of the different variables across multiple multitasking weeks (only weeks with Projects $>$ 1 were modeled). To perform the analysis, we first removed outliers. To capture developer-to-developer variability in the response ($FileTouches$), (e.g., some developers being naturally more productive than others), rather than assessing the contributions of specific developers, we fit a linear mixed-effects model with a random-effects term for the developer. All other variables were modeled as fixed effects. The linear mixed-effects models were used, as implemented in the functions \texttt{lmer} and \texttt{lmer.test} in R. Coefficients are considered important if they were statistically significant (p $<$ 0.05). Their effect sizes are obtained from \texttt{ANOVA} analyses. We evaluate our model’s fit using a marginal ($R^2_m$) and a conditional ($R^2_c$) coefficient of determination for generalized mixed-effects models, as implemented in the \texttt{MuMIn} package in R: $R^2_m$ describes the proportion of variance explained by the fixed effects alone; $R^2_c$ describes the proportion of variance explained by the fixed and random effects together. Table \ref{tbl:ghvsgl} presents summary statistics for our filtered data set compared to the statistics of the original study and, the results of the regression and comparison with the original research are shown in Table \ref{table:coefficients}.

\section{Results}\label{sec:results}

In this section, we present the results and the answers to our research questions.

\subsection{RQ1: What are the trends of, reasons for, and effects of multitasking and focus switching on developer productivity in GitLab?}\label{sec:rq1}

\subsubsection{Amount of Multitasking} 
\textbf{Do developers multitask?} To understand if developers multitask, we examined developers’ commit activity and also asked interviewed participants to report the number of tasks/projects they contribute to on an average day and week. From the repository analysis, we found that multitasking across projects over a week is not uncommon: developers contributed to multiple projects in 22.37\% of the developer-weeks in our dataset (or 3,279 out of 14,655), while OSS developers contribute in 37\% of the weeks. Industry developers work on average, 2.41 projects per week; OSS developers work an average of 2.57 projects per week. When we compare the daily multitasking results, we have similar behavior: while industry developers contribute to 1.49 on $AvgProjectsPerDay$ (min 1; max 6), OSS developers contribute to 1.44 (min 1; max 6). 

\textbf{Does within-day multitasking scale with the number of projects per week?}
In the original study, it was investigated whether contributing to more projects each week is also associated with contributing to more projects daily (therefore with more within-day focus switches), as one would expect. Our findings show that \textit{the industry developers do less context switches with a similar number of projects per week and per day that OSS developers}

\subsubsection{Reasons for Multitasking}

When interviewing the industry developers, we investigated reasons for working in multiples tasks or projects during the working day. Some of the reasons mentioned were support questions from other teams and issues from production environments that require the developers to stop what they are doing to understand the problem and to solve it. From the original study, developers mentioned interdependencies, personal interest, and social relationships. Also, typical responses were that they are an end-user of the software tool and want to fix bugs impacting it.

The dynamism of the work environment, multidisciplinary and social interaction were also mentioned as reasons to change focus: developers want to be part of different conversations, talks, and participate in meetings that are not directly linked to the projects they are currently allocated. One developer said: \textit{"You cannot sit at your desk the entire time, you need to talk to other people to have your work done"}. From the original study, developers stated interdependencies, personal interest, and social relationships as strong reasons for contributing to multiple projects. The need to change the focus, for the sake of working on something different, was also mentioned in the original study.

There are also mentions to focus missing, anxiety, and procrastination. A developer mentioned: \textit {"When you do not want to work and do not want to really solve something, you begin many tasks and do not finish anything."}

\begin{table*}[t!]
\center \small
\caption
{
Multitasking productivity model. GH (GitHub Data, Original Study): The response is log(LOCAdded) per week. \\$R^2_m = 0.35. R^2_c = 0.55$. GL (GitLab Data, Replication Study): The response is log(FileTouches) per week. $R^2_m = 0.22$. $R^2_c = 0.37$.
}
\label{table:coefficients}

\begin{tabular}
{
l D{)}{)}{20)5} D{.}{.}{5.5}  D{)}{)}{20)5}@{} D{.}{.}{5.5} 
}

\toprule

& \multicolumn{2}{c}{GH} & \multicolumn{2}{c}{GL} \\
\midrule
 & \multicolumn{1}{c}{ Coeffs (Errors)} & \multicolumn{1}{c}{ Sum Sq.} &  \multicolumn{1}{c}{ Coeffs (Errors)} & \multicolumn{1}{c}{ Sum Sq.}\\
\midrule
(Intercept)                                          & 0.069  \; (0.012)^{***} & -                & -0.039 \; (0.030)       & -              \\
GlobalTime                                           & -0.037 \; (0.005)^{***} & 15.71^{***}      & -0.010 \; (0.020)             & 0.16 \\
Projects                                             & 0.263 \; (0.006)^{***}  & 2566.87^{***}    & 0.217 \; (0.038)^{***}      & 20.07^{***} \\
Languages                                            & 0.549 \; (0.004)^{***}  & 11505.20^{***}   & 0.278 \; (0.023)^{***}      & 94.12^{***} \\
$S_\text{Focus}$                                     & -0.300 \; (0.004)^{***} & 2757.75^{***}    & -0.189 \; (0.025)^{***}     & 34.53^{***} \\
$S_\text{Language}$                                  & -0.231 \; (0.003)^{***} & 2354.39^{***}    & -0.423 \; (0.021)^{***}     & 244.35^{***} \\
AvgProjectsPerDay                                    & 0.046 \; (0.004)^{***}  & 215.29^{***}     & 0.022 \; (0.021)             & 0.67 \\
$S_\text{Switch}$                                    & 0.225 \; (0.004)^{***}  & 2255.09^{***}    & 0.206 \; (0.019)^{***}      & 72.60^{***} \\
Projects:$S_\text{Focus}$                            & 0.032 \; (0.003)^{***}  & 0.16^{***}       & 0.043 \; (0.016)^{**}       & 4.40^{**} \\
Languages:$S_\text{Language}$                        & -0.045 \; (0.002)^{***} & 345.961^{***}    & 0.014 \; (0.015)            & 0.57 \\
Projects:$S_\text{Language}$                         & -0.021 \; (0.003)^{***} & 68.66^{***}      & -0.013 \; (0.018)            & 0.30 \\
Projects:AvgProjectsPerDay                           & -0.024 \; (0.003)^{***} & 128.00^{***}     & -0.022 \; (0.015)            & 1.38 \\
$S_\text{Language}$:AvgProjectsPerDay                & 0.026 \; (0.003)^{***}  & 59.86^{***}      & 0.007 \; (0.018)            & 0.09 \\
Projects:$S_\text{Switch}$                           & -0.104 \; (0.003)^{***} & 404.80^{***}     & -0.081 \; (0.017)^{***}     & 14.62^{***} \\
AvgProjectsPerDay:$S_\text{Switch}$                  & 0.058 \; (0.003)^{***}  & 203.12^{***}     & 0.019 \; (0.017)            & 0.81 \\
$S_\text{Language}$:$S_\text{Switch}$                & -0.029 \; (0.003)^{***} & 35.99^{***}      & 0.015 \; (0.017)             & 0.47 \\
\midrule
\multicolumn{5}{l}{ GH: AIC = 171919; BIC =  172105; LogLik = -85939 ; Num. obs. = 78552; Num. groups: fUserID = 1193}\\ 
\multicolumn{5}{l} { GL: AIC = 8198; BIC =  8320; LogLik = -4079 ; Num. obs. = 3269; Num. groups: fUserID = 372} \\

\bottomrule

\multicolumn{5}{l}{\scriptsize{$^{***}p<0.001$, $^{**}p<0.01$, $^*p<0.05$}}\\
\end{tabular}
 
\end{table*}

\subsubsection{Productivity Effects}

\textbf{Is multitasking associated with more outputs produced per unit time?} In the original study, as a preliminary quantitative analysis, the authors found that more focus switches are associated with higher $LOCAdded$. In this replication, we turn to the multiple regression analysis, and from Table \ref{table:coefficients}, we can see the effects of the independent and control variables for the multitasking productivity model; the response is $log(FileTouches)$. Stars indicate the statistical significance. We can say that focusing more  (on fewer projects) and working more one day to the higher productivity, but there is no effect for higher multitasking ($AvgProjectsPerDay$).  We note the interaction between Projects and $S_{Switch}$: increasing the weekly total number of projects is associated with increases in outputs produced when developers do not follow too repetitive day-to-day patterns.  This suggests that taking on many prevent boredom and be beneficial.

\subsubsection{Perceived Impacts of Focus Switching}

The repository analysis is limited, so the interviews help to understand other impacts.

\noindent\textbf{Positive Impacts.} When talking about productivity, developers mentioned that small context change does not impact and does not disturb. For example: To run a script that it will take 15 seconds. During this time, developers mentioned that they could take a look at some other thing. These short changes of context do not impact too much, and do not disconnect them of the previous task. However, three or more minutes would be to much time. A developer said:

\textit{"I did not solve both tasks, and I feel that I left the first one behind."}.

In majority, developers said that it does not increase or decrease the time to delivery, but it delays. A lot of time is needed to rebuild the previous context, and something can be missed. However, multitasking can also make it possible to finish many little tasks at the same time. Or, it can block another task that needs more time to be completed. Most of the interviewed developers said that the code quality is not impacted by multitasking.

\noindent\textbf{Negative Impacts.} A single experienced developer said that multitasking is always negative. Two, believe that their productivity decreases. Others said that it is worst for coding tasks or when it takes to much time to come back to the previous task (as mentioned previously).When asked about how does the length of the switch affects productivity and code quality, most of the developers said that it is proportional to how much context is being switched: as the bigger the context switch and the longer the interruption, the harder it is to come back. Some comments were:

\textit{"We work to deliver for different tasks, however, less from each one."} 

\textit{"It breaks the flow of work, I cannot focus in only one task. It impacts my line of sight."} 

About code quality, only one thinks that it decreases (the most junior developer interviewed). One developer said that it is hard to solve hard problems with too many interruptions. Developers surveyed in the original study believe that switching context can drag productivity and introduce more bugs.

\noindent\textbf{Happiness versus Stress Levels.} We asked developers how much their happiness and stress levels are affected by multitasking, a question that was not in the original study. In general, developers think that happiness is impacted in the medium to long term. If multitasking frequently happens, daily, for example, they begin to feel not productive anymore, and it affects happiness:

\textit{"Happiness decrease. I feel tired at the end of the day."} 

\textit{"The days I do not use a time management technique, I feel tired, and I have a feeling that I did not reach my goal."} 

All developers said that stress levels increase at some point and, it becomes frustrating and impacts the process and personal organization. However, a developer mentioned that he uses stress and a high number of tasks to motivate himself:

\textit{"In the worst case, the stress level increase. Happiness can vary greatly. When you can solve everything, the high-stress level becomes happiness. I think working in a large number of tasks can make me happier. I use this as a positive mental model, but thinking out loud, it seems not so good."}

\subsection{RQ2: Are there limits on multitasking (and what are they) before productivity is impacted?}

\subsubsection{Quantitative Analysis}

Table \ref{table:coefficients} yields interesting two-way interactions. The model of the original study fits the replication study data acceptably if we compare the $R^2_m$ values directly. However, it is important to consider the following: 1) the original model used more data 78,000 rows compared to only 3,000 rows of the replication - the more the volume the data we have, the more effects you can detect (and more coefficients will likely be statistically significant in the model); 2) the original model used LOCAdded/week as the outcome variable, compared to FileTouches/week in the replication, so it might be the case that LOC models fit the data better than FileTouches, even if the two are highly correlated.

The signs of the main coefficients ($AvgProjectsPerDay$, $S_{Focus}$, $S_{Switch}$) are all similar between the two models (original and replication) except for $AvgProjectsPerDay$ which has no statistically significant effect now. So, we can say that focusing more (on fewer projects) and working more repetitively from one day to the next is associated with higher productivity, but there is no effect for higher multitasking ($AvgProjectsPerDay$). We can also note the interaction between Projects and $S_{Switch}$: increasing the weekly total number of projects is associated with increases in outputs produced when developers do not follow too repetitive day-to-day patterns. It suggests that taking on many projects can prevent boredom and be beneficial; however, if switching between them becomes too repetitive, productivity decreases.

\subsubsection{Developer Perceptions on Limits}

When asked about perception on limits, developers mentioned that it is proportional to how much context is being switched: as bigger the context and bigger the interruption, it is worst to come back. Also, they perceive that limits were crossed when stress scales, as mentioned, for example, by a developer: 

\textit{"You are giving support to fix an issue, so another problem appears and then some bureaucracy, and on, and on."} 

When the work becomes boring (the same kind of task regularly), they reported that multitasking becomes a way to deal with that.

\subsection{Insights From Questions of the Interview }

Additionally to the original research questions, the interview brought to the table some others qualitative insights we share in this Section.

We asked the industry developers which kind of switches could be avoided or rescheduled. Some of the answers were: Change of priorities that come from external sources; reducing support calls by introducing better diagnosis tools for production incidents; distractions from non silent open environments; to protect developers from external interruptions (for example, better filtering coming from product owners and scrum masters)

We also asked the typical triggers for switching between tasks, and change of business rules; lack of knowledge in a specific subject; and communication issues were some of the answers.

Developers answered about their typical resume strategies for returning to a previous task. Some do not have strategy at all, just let the source code opened where the developer was working before. Other answers were: to let code non compiling where the developer was working on; code annotations; to do lists; Git features as Git Stash to come back to previous state; check the differences between Git and the local state of the code.

\section{Discussion Across Studies}\label{sec:comparison}

Through this section, we summarise and compare the main results from both studies. OSS developers use to multitask in more developer weeks than industry developers (37\% versus 22.37\%); however, the $S_{Focus}$ (diversity on focus switch) are similar 0.92 to OSS developers and, 0.94 to industry developers.

There are some common reasons why OSS and industry developers do multitasking and work in multiple projects: dependencies, personal interests, and social relationships (to gain reputation in the OSS community or the company they work on, for the industry). Developers working for industry mentioned an additional reason for multitasking: bug fixing (mainly in production environments). From human and social aspects, industry developers said that multitasking sometimes is generated by focus missing, anxiety, and procrastination.

There were also comments about the perceived impacts of focus switching. OSS developers mention as positive the possibility to contribute with multiple projects increases, the chances of each project succeed. Also, they said that working in multiple projects makes it possible to resolve more issues. For industry developers, short periods of focus switching do not impact or disturb their work. Also, they reported that multitasking makes it possible to finish many little tasks at the same time. As negative impacts, OSS developers do not believe that multitasking has a positive impact on code quality as well as context switching. The industry developers also believe that productivity decreases proportionally to how much context is being switched; however, only a few mentioned that code quality decreases. As more significant the context being switched and the interruption, harder is to come back. A recent study from Duncan et al. \cite{Brumby2019} says that there is evidence to support the idea that following an interruption, people fail to remember what they were doing in a task before being interrupted. Also, there is a link between how quickly a task is resumed and the likelihood that an error is made. The same study says that interruption researchers have generally considered a longer resumption lag to be a bad thing — reflecting time needlessly wasted following an interruption. 

An analysis comparing happiness versus stress level, not present in the original study, shows that happiness of industry developers is impacted in medium to long term, and they do not feel productive anymore. Stress levels increase at some point, leaving developers frustrated.

OSS and industry developers have different perceptions of multitasking limits. While OSS Developers want to increase the number of projects they contribute, industry developers use multitasking to deal with tedious tasks.

\section{Threats to validity}\label{sec:threats}

The assessment of Threats to Validity (TTVs) is critical to secure the quality of empirical studies in Software Engineering \cite{Zhou2016}.  Our threats to validity are mentioned below, based on categories of threats mentioned by Wohlin et al. \cite{Wohlin2012}
\begin{itemize}

\item \textit{\textbf{Reliability}}: This aspect is concerned with to what extent the data and the analysis are dependent on the specific researchers. We had a few numbers of interviews, and they are all performed by only one person. Two risks are identified here: unclear coding of collected data and a possible risk of researcher bias.

% \item \textbf{\textit{Internal validity}}: Threats to internal validity are influences that can affect the independent variable with respect to causality, without the researcher's knowledge. Thus they threaten the conclusion about a possible causal relationship between treatment and outcome. For this replication,  we used the same regression analysis model of the original study without changing, so we are considering this as a risk to the causal conclusion. Also, the original study conceptualizes as a project, repositories that are grouped when owned by the same GitHub user, or having precisely the same name into sets of repositories. This was already presented as a threat. For the replication, once the data came from GitLab of the company, repositories were already under a project, and we considered the data organization as it is.

% [Bogdan comment] We can't really claim casualty with our study design (because it's not an experiment with random assignment to either the treatment or the control group). Which is why we rely on theory to describe a potential causal link, and why we statistically control for covariates in the multiple regression model, that random assignment in an experiment would otherwise control for.

\item \textit{\textbf{External validity}}: The concern of this aspect is whether the results can be generalized. In the original paper, the projects extracted were based on data from GitHub, which limited the generalizability to that context. In this replication work, we enlarged the sample and enriched the discussion about multitasking using data from a technology company. However, we analyzed data of only one company, and we had a few interviews, but we see that as a step forward to more generability.
\end{itemize}
 
\section{Research Agenda}\label{sec:agenda}

The results of this replication study brought emerging significant insights on multitasking and focus switching across projects when comparing OSS developers and developers that dedicate most of their time to industry; however, we want to perform additional analysis through the industrial data and collect additional insights than the ones obtained in the original study. We understand that we will be opening some interesting avenues for future research.

\textbf{Further investigation on different companies.} This study compares the multitasking work of OSS developers in GitHub with industry developers from one company only. It would be essential to gather the repository information of other companies (that would like to share their data anonymously) to have different samples and better understand the impact of the cultural environment on the results. It would help to build a solid base of knowledge, reducing uncertainty about theories, methods, tools, etc., helping to strengthen the evidence.

\textbf{Further investigation on different genders.} The multitasking behavior impacts differently when considering different genders? From the data gathered from both OSS and industry, it is possible to infer gender using different algorithms available. Future work can investigate if multitask is different for men and women and consider a different kind of qualitative research to support the findings.

\textbf{Further investigation on team's self-evaluation.} Another point to be evaluated is how much multitasking impacts the self-evaluation of the industry development teams. The company analyzed runs, for every development cycle (sprints of different sizes), a survey built over five qualitative variables: Goals and Planning, Productivity and Delivery, Technical assumptions and Quality, Process and Ceremonies, and Collaboration across Teams. The objective is to be a tool for continuous improvement, so the survey is usually answered after the retrospective of the development cycle when the teams discuss how it was. There is a possibility to include in the discussions of the teams, points about focus changing, and the impact on the work.

\section{Conclusions}\label{sec:conclusions}

Developers mentioned that small context change did not impact and gave a number: three minutes. More than that brings a feeling of left the previous task behind. It is proportional to how much context is being switched: as bigger the context and bigger the interruption, it is worst to come back. Hard problems become harder to solve when interruptions are in place. Borst et al. \cite{Borst2015} have shown that the disruptiveness of interruptions is for a significant part determined by three factors: interruption duration, interrupting-task complexity, and moment of interruption corroborating with the findings of both studies.

% For industry developers, focusing more (on fewer projects) and working more repetitively from one day to the next is associated with higher productivity, but there is no effect for higher multitasking (AvgProjectsPerDay). Also, the interaction between Projects and $S_{Switch}$ suggests that taking on many projects can prevent boredom and be beneficial; however, if switching between them becomes too repetitive, productivity decreases due to increasing the weekly total number of projects is associated with increases in outputs produced, when developers don’t follow too repetitive day-to-day patterns.

We also found out that industry developers multitask as much as OSS developers. There are commons reasons for both groups to multitask: dependencies, personal interests, and social relationships. Dependencies are commonly related to code they need from other teams or developers prior synchronize work to continue (e.g., a front end development task that depends on an API developed by others). 

Personal interests are related to technologies they want to learn, problems that require knowledge they do not have. Social relationships are related to how well they go across the company, collaborate with other teams, projects, and their reputation coming from that. 

The need to change the focus, for the sake of working on something different, was also mentioned by both groups. 
Industry developers also mentioned focus missing, anxiety, and procrastination as reasons for multitaasking.

\section*{Acknowledgment}
This project is partially funded by FAPERGS, project 17/2551-0001/205-4.

\bibliographystyle{ACM-Reference-Format}
\bibliography{sample-sigconf}

\end{document}